# Elimination of the acoustoelectric domain and increasing of the emission intensity by means of shunting the lateral current in the InGaAs/GaAs heterostructures

P.A. Belevskii, M.N. Vinoslavskii[1], O.S. Pylypchuk

*Institute of Physics, National Academy of Sciences of Ukraine, 46, Nauky Avenue, 03028 Kyiv, Ukraine*

**Abstract**

There has been experimentally implemented a technique of neutralization of an acoustoelectric domain under the conditions of the lateral transport of charge carriers in strong electric fields in the multilayer InGaAs/GaAs heterostructures with quantum wells. The technique is implemented by means of deposition of a shunting semi opaque silver film on the heterostructure surface between the ohmic contacts. In absence of shunting the domain appearance leads to current decrease, current oscillations and also to a strong decrease of the band-to-band emission during a voltage pulse applied to the sample. The shunting eliminated the current decrease and enabled it to strongly enlarge the band-to-band emission intensity.

## I. Introduction

Currently a significant interest is kept to investigations of effects related to the lateral (along heterolayers) transport of the charge carriers in strong electric fields in multilayer heterostructures (HS) with quantum wells (QW). These are transport non-linear effects and current instabilities in the lateral electric field caused by spatial interwell transitions of charge carriers due to heating up by the field [1,2], an impact of a barrier width between double QW InGaAs/GaAs on the bipolar transport and terahertz emission in the lateral electric field [4]. In such structures one succeeded to enlarge the carriers lifetime by several orders of magnitude and the drift length up to hundreds μm in HS with the InGaN/GaN QWs [5] or up to several mm in HS with the InGaAs/GaAs QWs [6]. Such effect arises due to spatial separation of electrons and holes among quantum wells in a heating electric field. The band-to-band and intraband terahertz emission in the InGaAs/GaAs HS was observed in the lateral electric field from extended (3 through 4 mm) structure surface [7,8]. Such property of similar HS implies a potential possibility to create emission sources on their base in near-infrared or terahertz wavelength ranges.

Along with that, a strong acoustoelectric instability arises in the course of investigating the lateral transport of the charge carriers under strong electric fields in the elongated HS with QWs based on the piezoelectric semiconductors like GaAs. This instability was considered theoretically [12,13], extensively explored in such bulk piezoelectric semiconductors as CdS, GaAs, InSb, Te, as well as in heterostructures with QWs based on these semiconductors.

---

[1] Corresponding author: mvinos@ukr.net

It arises when the electron drift velocity exceeds the sound velocity, and the intensive acoustic phonons emission begins in the dedicated forward direction. This results in enhancement of the sound waves in a narrow frequency range which propagate in the electron drift direction. These waves create a transverse acoustic wave which modulates the semiconductor energy band edges and creates a periodic piezoelectric field. Interaction of charge carriers with this field results in their capture by the piezo wells of the wave [19 - 21] and their retarding. As a result, there occurs redistribution of the electric field in the crystal and creation of a very narrow spatial region of the sufficiently strong electric field and high phonon density - acoustoelectric domain (AED). These domains arise in the region of the enhanced acoustic noise near the cathode and propagate to the anode with the sound velocity. They present themselves an narrow pack (~100 μm) consisting of a set of piezoelectric wells filled with electrons. Creation of the acoustoelectric domain on the background of the spatially distributed piezoelectric wave occurs during a short "incubation period". During this period the current through the crystal is almost unchanged. At the incubation period end the further domain creation and its movement to the anode contact occurs, which is accompanied by the current decrease.

The evolution of acoustoelectric instability was used to explain the current oscillations under strong electric fields in the AlGaAs/GaAs HS with QWs [10] and InGaAs/GaAs [11], as well as nGaN [23]. The acoustoelectric instability evolution in the InGaAs/GaAs HS with QWs is accompanied by the acoustic wave generation and appearance of the strong field acoustoelectric domains (AED) [12-14]. The AED creation leads to the current decrease and strong (by an order of magnitude and even more) decrease of the band-to-band or terahertz emission [7,8].

Emergence of moving AEDs is used in a number of papers for different applied investigations. So, in Ref [23] the emergence of moving AED in the n-GaN HS was used to modulate the terahertz emission from an aside source. The THz wave modulation was used for various applications, including wireless communication, quantum electronics, spectroscopic applications and visualization [27,28]. There are proposed THz modulation by a non-uniform density of free charge carriers in a semiconductor [29], two-dimensional plasmons [30], intraband absorption in the two-dimensional materials [31,32], and two-dimensional optoelectronics. The AED generation was used to shift the absorption edge, the light polarization plane rotation, to stimulate light emission and the Brillouin scattering in the visible spectrum range, as well as weakening the microwave emission transfer.

Investigation of the spectral dependence and polarization anisotropy of the visible light transmission modulation by the AE domain at the semiconductor intrinsic absorption edge has shown the modulation signals to be caused by the Franz-Keldysh effect arising due to a strong electric field resulting from the AE potential fluctuation inside the domain [24-26].

Emergence of the acoustoelectric domains restricts creation of the light emitting devices in such structures in spite the possibility for electrons and holes to fill large lengths (~ several mm) in them and obtain a large emitting surface.

A question arises whether one can suppress emergence of acoustoelectric instability in such structures and observe a non-weakened emission. One proposes, for example, in theoretical Refs [15,16] to deposit a shunting metallic film along the side surface of the QWs stack between the top and bottom contact layers in the multilayer HS destined for the vertical current transport and obtaining the terahertz emission [17]. This is destined to even the vertical distribution of the electric field along the QWs stack and suppress emergence of possible electric instabilities.

We propose to carry out such shunting along the QW layers between the end contacts in the multilayer HS with QWs destined for the lateral current transport with the emission output through HS surface.

## II. Samples and experimental technique.

The impact of the shunting metallic layer on the lateral conduction and band-to-band electroluminescence in the multilayer n-$In_xGaAs_{1-x}$/GaAs (x = 0.1) heterostructures with quantum wells (QWs) has been studied. The structures consist of 20 single quantum wells with a width of 20 nm, which are divided by the 90 nm wide barriers. The GaAs layers are doped by Si with the concentration of $n_s$ = 4.1·10$^{11}$ cm$^{-2}$ per period. The buffer, spacers between QWs and cap layer were 100 nm wide. The heterostructures were grown by the MOVPE method on the substrate of the semi-isolating GaAs (001) with the 400 μm thickness.

The rectangular shape samples of the 5 through 6 mm length and 2 though 3 mm width were cleaved from the heterostructure wafers (Fig.1). The ohmic contacts of the In, ~1 mm wide strips, were deposited and fused in the argon ambient at 430 C at both sides. The distance between contacts was 3 though 4mm. The rectangular voltage pulses of the 20 through 300 V amplitude were applied to the contacts along the [110] direction. The pulse duration was restricted by the established current value and was about 5 μs. The electric field dependences of the current and band-to-band IR emission intensity in the direction perpendicular to layers plane at QWs side have been measured (Fig.1). The band-to-band emission emerged in the samples with the In ohmic contacts at strong fields (>100 V/cm) due to interband ionization on the tips of the impurity shunts in the anode contact [18]. We measured both the integral band-to-band emission intensity from the whole sample surface and distribution of this emission along the current direction between contacts with use of the scanning gap of the 0.3 mm width. The photoelectron multiplier with the AgOcs photocatode (The spectral range in 0.4 through 1.2 μm) was used as a detector. The experiments were carried out at 77 K.

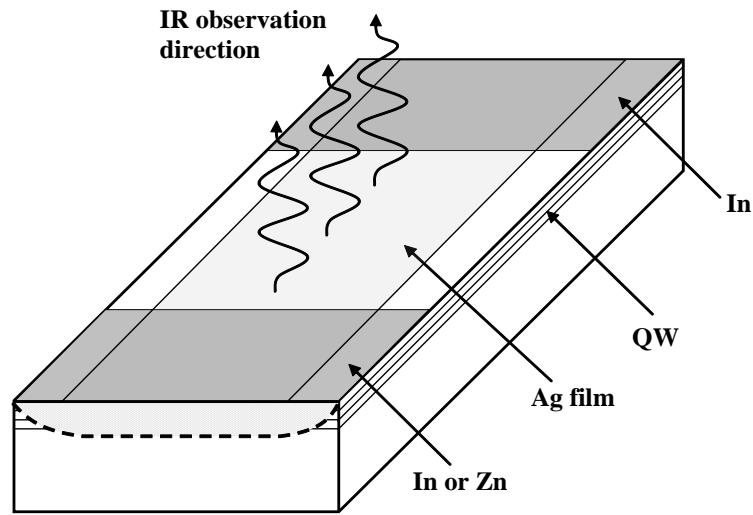

**Fig.1.** The sample sketch of the HS with QWs and the deposited shunting silver film.

The experimental peculiarity consists in that after carrying out the indicated measurements the thin semi opaque silver film was deposited in vacuum on the sample surface at side of QW and contacts (Fig.1). The film thickness was selected so to make the film resistance about by an order of magnitude larger as compared to the sample resistance at 77 K. The film served as a shunt for the lateral current along QW layers in order to smooth emerging local field non-homogeneities like, for example, in the case of emergence of the acoustoelectric domains [14]. After deposition the measurements circle repeated, and the comparison of the experimental results was carried out for the samples with and without a deposited film. There were carried out also experiments with different configurations of the deposited shunting film (at the substrate side with coming on the end contacts, side planes od HS, on the surfaces though without coming on the one of contacts). However, these attempts did not give a desirable result of shunting.

### III. Experimental results and discussion

#### A. Impact of the acoustoelectric domain on the current characteristics and band-to-band emission in the non-shunted InGaAs/GaAs heterostructures with QWs

The current and band-to-band emission oscilloscope waveforms under application of the rectangular voltage pulses with different amplitude and ~5 µs duration obtained for the single well heterostructure with the ohmic contacts and without shunting are shown in Fig.2. It is seen that the oscilloscope waveforms of current J and emission intensity IR weakly increase in the pulse top in the weak field (Fig.2a). This is caused by the drifting filling the sample by holes from the anode contact to the cathode. In the case of a strong field (Fig.2c) the current during the pulse beginning in the incubation part, $J_{inc}$, with duration of 100 through 400 ns sharply grows up to the maximum magnitude, and further one observes a sharp current drop and emission intensity decrease. This current drop from the magnitude of $J_{inc}$ down to the minimum value is accompanied by decaying oscillations [14] induced by emergence and movement of the acoustoelectric domain (AED). The IR emission oscilloscope waveform has also a narrow maximum in the incubation period and a subsequent sharp drop (by up to 20 times) (Fig.2c).

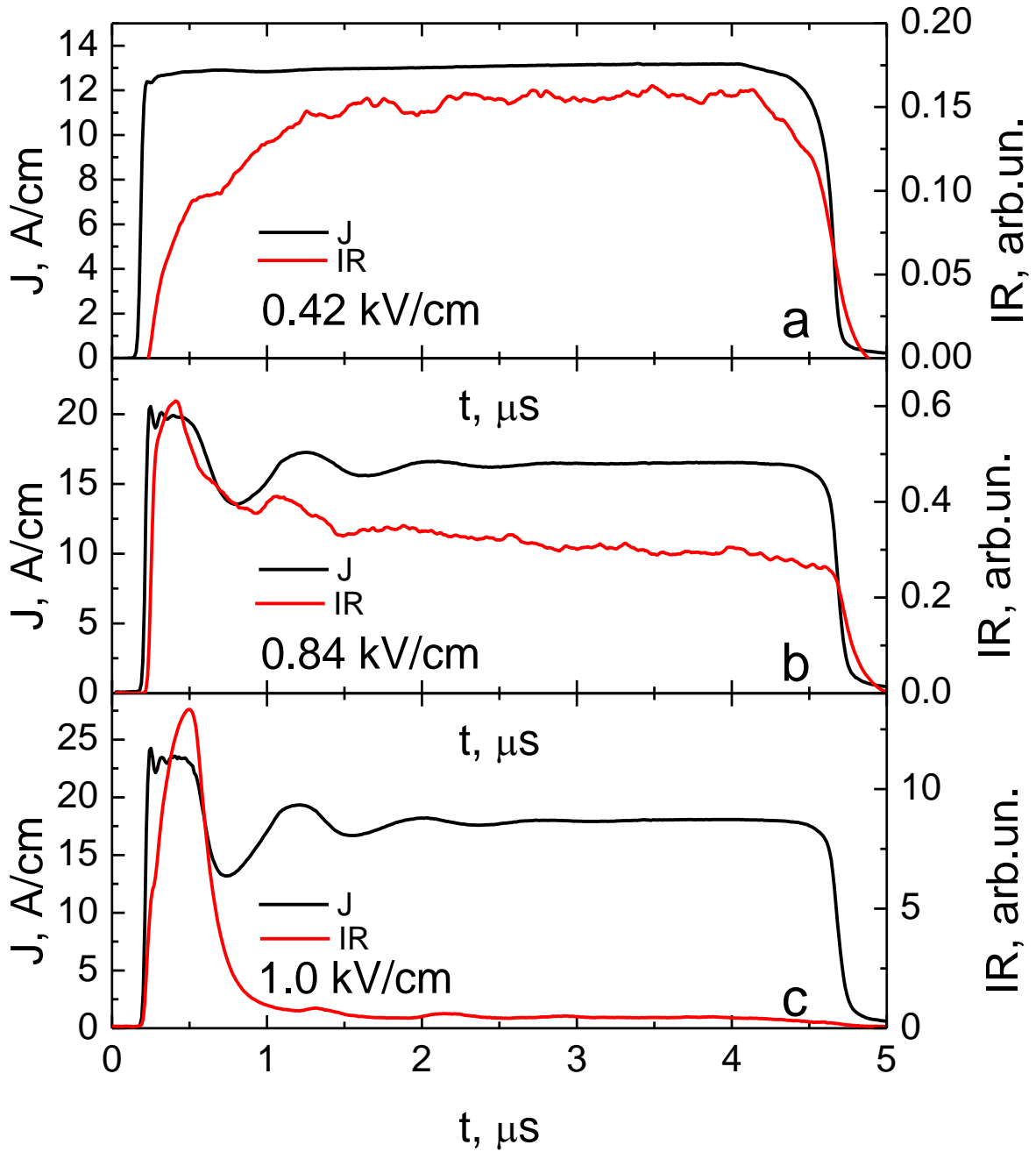

**Fig. 2.** The oscilloscope waveforms of currents (J) and integral band-to-band emission (IR) in the InGaAs/GaAs heterostructures with the ohmic (In) contacts without the shunting metallic film at different heating electric fields (a, b, c).

The field dependences of current (VCCh) and band-to-band emission (IR) for the sample without the shunting metallic film are shown in Fig.3. Inset shows trapping of electrons and holes into the piezo wells created by the electric field of the acoustic wave modulating the semiconductor band edges. This picture corresponds to the incubation part $J_{inc}$ of the field dependence of the current (VCCh) yet before the domain emergence. Also, these dependences of the band-to-band emission (IR) are shown in Fig.6 for comparison with such dependences in the sample with shunting. The field dependences VCCh and IR are shown for two times: at the beginning and the end of the current pulse. In the case of the non-shunted sample the field dependence of the current (VCCh) in a not high electric field (E<0.5 kV/cm) and corresponding to the pulse end goes somewhat higher than the curve VCCh corresponding to the beginning of the pulse. The dependence of IR(E) corresponding to this part of VCCh at the pulse end goes also considerably higher than the IR(E) in the beginning of the pulse (Fig.3). As was mentioned above, this is caused by drifting filling of the sample by the carriers and growth of the J and IR pulses.

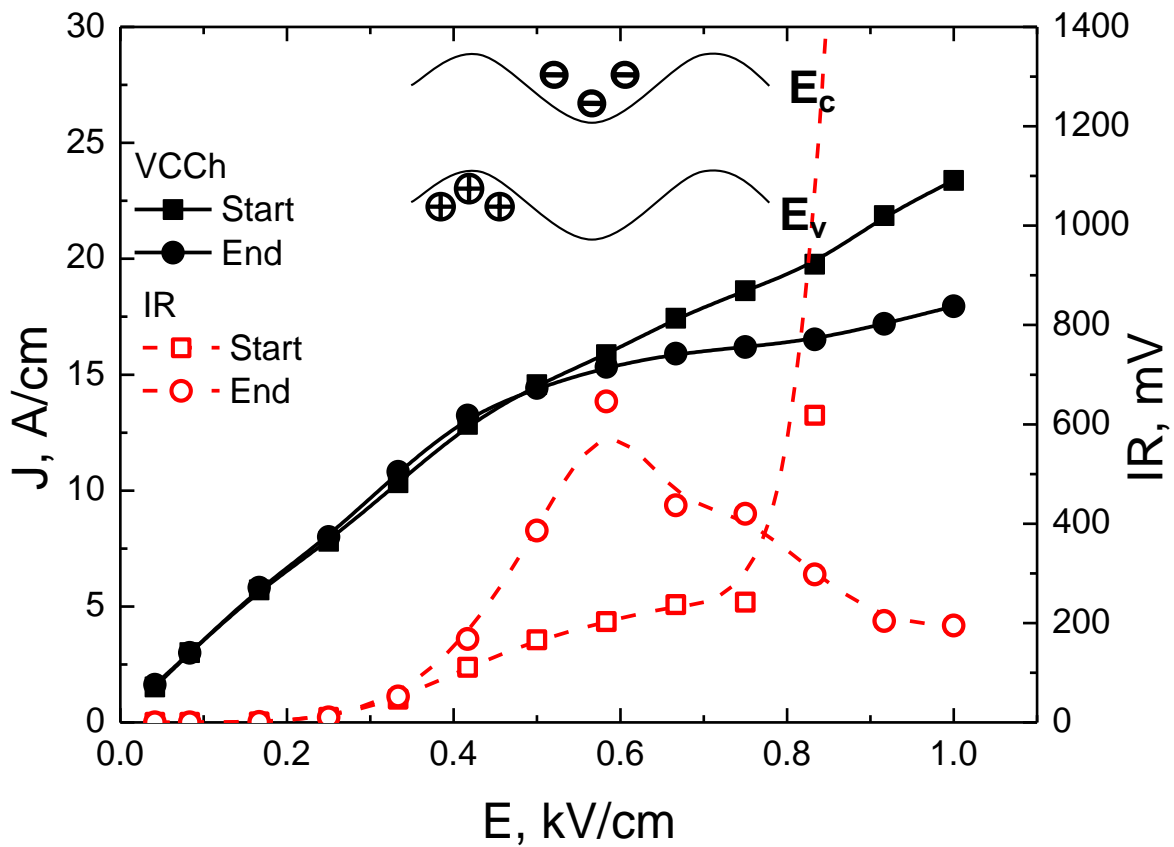

**Fig. 3.** The field dependences of the current (VCCh) and band-to-band emission (IR) in the heterostructure at the beginning (Start) and the end (End) of the pulse for the samples without the shunting metallic film at different heating electric fields. Inset: trapping of the photogenerated charge carriers into the piezo wells created by the acoustoelectric wave as illustrated in Refs 19.

In the next part of the VCCh curve (E>0.5 kV/cm) for the not shunted sample one observes its slope decrease and division into two branches – at the beginning and the end of the current pulse. At that, the VCCh at the beginning of the pulse corresponds already to the so-called time of the acoustic domain incubation (Fig.2 b, c) and goes above the VCCh curve at the pulse end. The domain begins to form itself after this point (by field and time). It is accompanied by a slope decrease of the IR(E) dependence because of trapping electrons into the domain piezo wells. At that, the domain already has formed itself at the current pulse end and localized near the anode contact (Fig.2 b, c) [14]. Some decrease of the VCCh curve slope at the beginning of the pulse in the $J_{inc}$ part after its bifurcation is caused by a part trapping electrons during the incubation period (before domain emergence) into piezo wells created by the acoustic wave field (Fig.3, inset).

The sublinear growth of the VCCh curve for the not shunted sample at the pulse end reflets creation of the acoustoelectric domain in the fields $E \geq 0.5$ kV/cm which strongly restricts the current in the sample (Fig.3). The electron mobility in our samples measured by the Van der Pauw method at 77 K in the weak fields is about 5000 cm$^2$/V·s. Hence, the electrons achieve the sound velocity, which in InGaAs is about $3.5 \cdot 10^5$ cm/s, already in the field of $E \leq 100$ V/cm. However, enhancement of the acoustic waves in our case occurs only in the narrow QW layers (totally 0.4 µm). At that, the acoustic wave apparently to overlap the whole sample bulk together with the substrate (400 µm). Therefore, the acoustoelectric domains creation occurs at significantly higher fields $E \geq 0.5$ kV/cm (Fig.3).

The intensity of the band-to-band emission $IR_{inc}$ from the samples with the ohmic contacts and without the shunting metallic film is small at fields $E < 0.8$ kV/cm. It may be explained, first, by a still insufficient field for the interband ionization at the anode contact. Secondly, it may be affected by emergence of the coherent acoustic waves in the sample which modulate the potential energy of the crystal [19,20]. The so-called "piezo wells" emerge in result in the sample, which trap electrons. And only in the fields of $E > 0.8$ kV/cm the emission begins strongly to increase (Fig.3) that may be related to the large rate of carrier generation at the anode contact, as well as output of electrons out of piezo wells in a strong electric field. However, this strong growth is characteristic only for the incubation period at the beginning of the pulse (Fig.2, time $t_1$) which has small duration (about 200 ns). As a result of the AED emergence and its localization near the anode contact the band-to-band emission sharply drops vs time (Fig.2b, c) and its sharp growth (at the current pulse end) is not observed (Fig.3). It is already related to the trapping electrons into piezo wells in the AED region. One should note that the piezo waves generated of the near cathode noise near the cathode contact [12] move toward the anode contact. They can trap only electrons, because the holes move in the opposite direction from the anode to the cathode. The reflection of the acoustic wave from the anode and its movement in the direction coinciding with hole drift to the cathode leads already to trapping holes into the wave piezo wells and releasing electrons. Namely this results in the decaying current oscillations (Fig.2b,c). In this case the spatial separation of electrons and holes occurs. It enables electrons and holes to drift over the whole sample length. The electrons in sufficiently strong field during the incubation period can leave the piezo wells due to heating and recombine with holes more efficiently. It results in a sharp emission intensity growth (Fig.3, curve $IR_{start}$).

Our situation differs from the papers in which the electrons and holes are generated pointwise by optic means in heterostructures with QWs by an independent light source and the acoustic waves also independently are generated in a heterostructure. In this case the acoustic waves may trap into piezo wells both electrons and holes (Fig.3, inset) and convey them over longer lengths in direction of a wave movement [19,20].

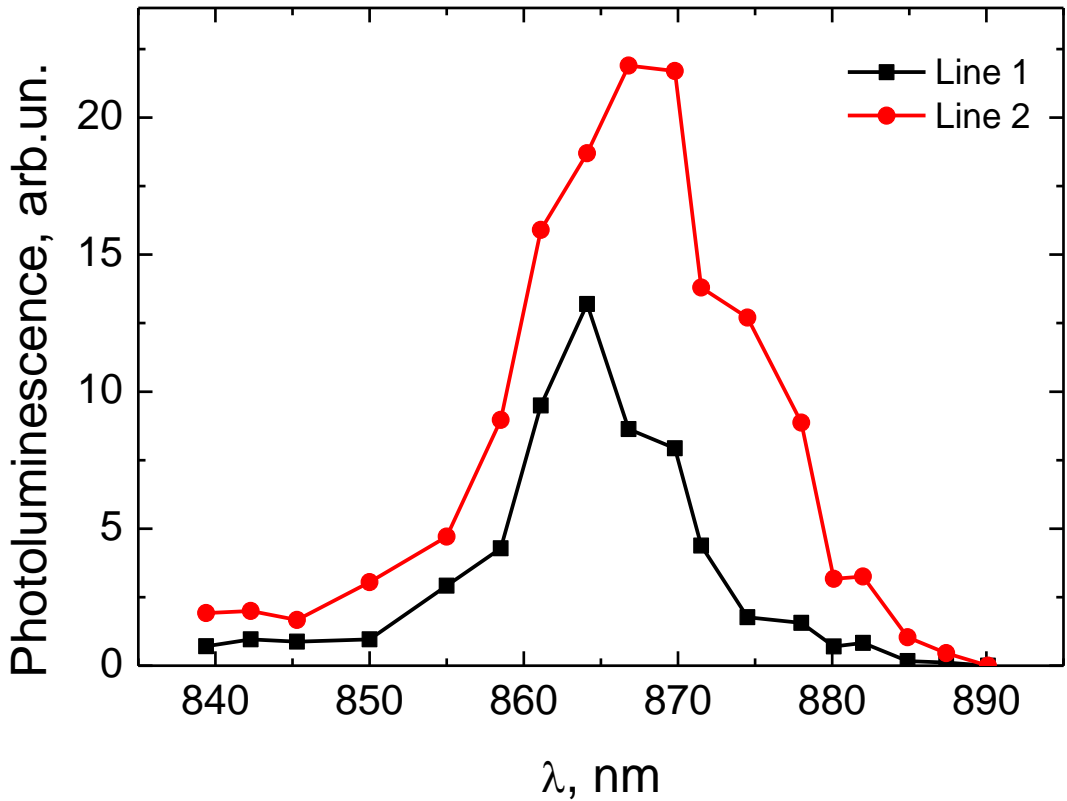

**Fig. 4.** The photoluminescence spectrum of the sample without shunting in absence of the field (line 1) and band-to-band electroluminescence in the electric field (line 2) at the beginning of the pulse. $E_1 = 0$ (line1), $E_2 = 1$ kV/cm (line 2).

The following fact may also be evidence of trapping electrons by the piezo waves. The spectral peak of electroluminescence in the electric field and in the beginning of the current pulse shifts into the region of less energy (longer wave lengths) as compared to the photoluminescence peak in absence of the electric field (Fig.4) (the Franz-Keldysh effect) [22]. The spatial separation of electrons and holes causes their direct recombination in the momentum space and indirect one in the real space goes.

### B. Impact o the shunting metallic film on the current characteristics and band-to-band emission in the InGaAs/GaAs heterostructures with QWs

The oscilloscope waveforms of the current and band-to-band emission due to action of the rectangular voltage pulses of different amplitude and duration of ~5 μs obtained for the heterostructure with ohmic contacts and shunting semi opaque metallic film are shown in Fig.5 a, b, c. One observes a weak increase of the current vs time in the current J pulse top region (Fig.5 a, b, c). This growth accelerates with an increasing field and then saturates. Such behavior is characteristic for injection of carriers from the anode contact. Since the current J increases during the voltage pulse it evidences of the AED absence in the shunted sample.

The IR emission intensity of the shunted sample in the established regime at the pulse end exceeds that in the not shunted sample by more than an order of magnitude. The IR emission signal at the fields E≤200 V/cm increases (Fig.5a) and at fields E≥500 D/cm slowly decreases during the whole pulse (Fig.5b, c). It is obvious that this decrease is related to the AED emergence. Also, we do not relate this IR emission signal decrease with the Joule heating of HS during the current pulse, since at such a long pulse duration (~60 μs) the IR signal after 10 μs in fact saturates (Fig.5c, inset). We can relate this emission decrease to accumulation of the acoustic phonons, generation of the piezoacoustic waves and partial trapping electrons in them.

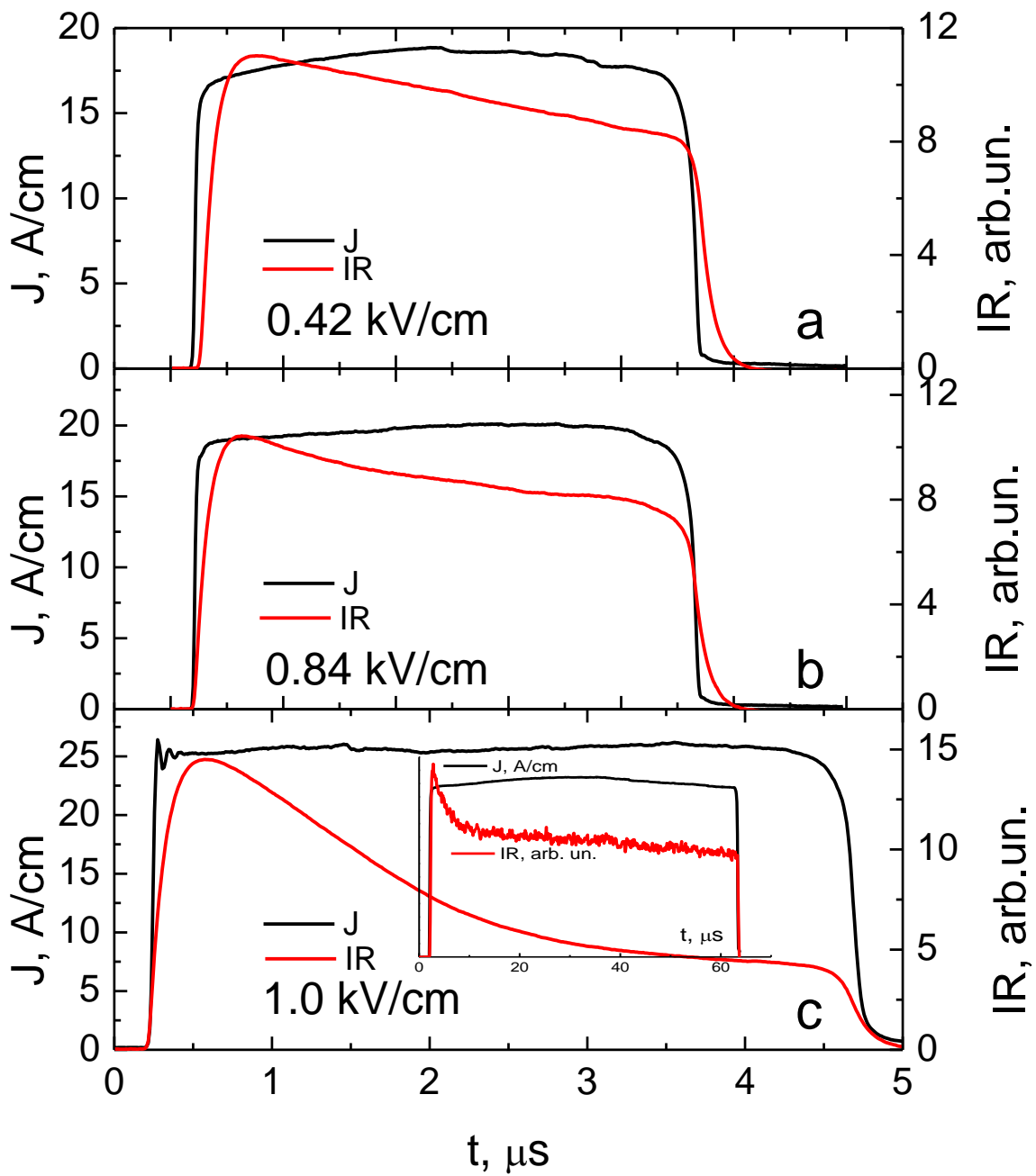

**Fig. 5.** The oscilloscope waveforms of the current (J) and integral band-to-band emission (IR) in the InGaAs/GaAs heterostructure with the ohmic (In) contacts with presence of the shunting film (a, b, c) at different heating electric fields.

In the case of the shunted sample (in contrast to the not shunted one) the VCCh curve at the pulse end goes above the VCCh curve at the beginning of the pulse in the whole range of fields. This discrepancy is considerably large than for the not shunted sample at the initial part of VCCh. Such behavior of VCCh may be related to a strong shunting impact on aligning the field distribution between the anode and cathode contacts. At that the near-cathode field, where an AED usually emerges, decreases and the near-anode field increases. It results in a stronger generation of the electron-hole pairs in that, a large current growth and considerably larger (by two orders of magnitude) growth of the band-to-band emission in the fields E<0.8 kV/cm (Fig.6) as compared to the not shunted sample.

Deposition of the shunting metallic film on the sample cardinally changes the field dependences of the band-to-band emission intensity. The intensity growth begins already in the fields of E~0.08 kV/cm (Fig.6). Then at field about of 0.3 kV/cm one observes saturation and further at E≥0.6 kV/cm the intensity weakly increases with a growing field at the beginning of the pulse and somewhat decreases at the pulse end. At E=0.3 kV/cm the emission intensity in the shunted sample is by two orders of magnitude higher than in the case of no shunting. One explains this by absence of the domain piezo wells in such a sample and trapping carriers in them.

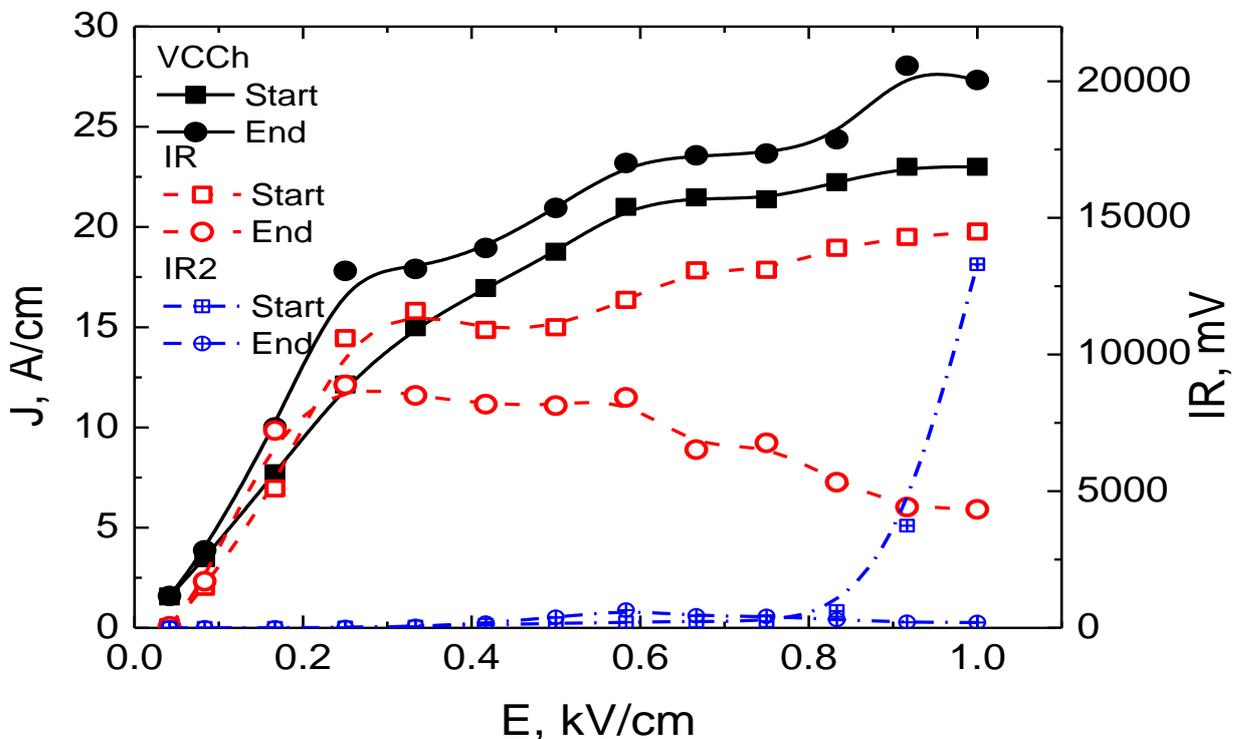

**Fig. 6.** The field dependences of the current (VCCh) and band-to-band emission (IR) in the heterostructure in the beginning (Start) and at the end (End) of the pulse for the sample with the shunting film at different heating electric fields. For comparison the band-to-band emission of the not-shunted sample is shown (IR2).

The current values though the shunted and not-shunted samples in the small fields E≤40 V/cm do not in fact differ from each other, since, as mentioned above, the shunting film resistance is higher by an order of magnitude than that of the sample. However, in strong fields the currents through the shunted sample exceed those through the not-shunted sample by about 1.5 times (see Fig.3 and 6).

The sublinear behavior of the VCCh curve in the beginning of the pulse in the case of the shunted sample (Fig.6) is similar to that the VCCh behavior in the incubation period for the not-shunted sample (Fig.3). Such similarity apparently has the same origin related to decrease of the carrier mobility because of an increase of the acoustic scattering, when the electrons achieve the sound velocity in GaAs and begin efficient generating acoustic phonons. It occurs independently of whether shunting is made or not. However, the shunting film on the sample surface makes alignment (shorting) of arising nonuniformity of the potential along the current, similar to that as was proposed in [15,16]. Therefore, the strong field acoustoelectric domain does not emerge in the shunted sample. A decrease of the VCCh curve slope during the incubation period in the not-shunted sample as well as that at the beginning of the pulse in the shunted sample ($VCCh_{start}$) apparently is affected by the partial trapping of electrons by the piezo waves emerging in the electric field yet before formation of the acoustoelectric domain. There is possible also heating of electrons in QW before output in the GaAs barrier. This causes increasing impurity scattering in the long impurity well in the barrier. Besides, the slope decrease of the $VCCh_{start}$ may be affected by heating the part of electrons up to the second size quantization level and localization of the electron distribution function in QW near barrier walls. In consequence there increases their scattering by interfaces while the maximum of the distribution function of electrons at the first size quantization level is in center of QW.

Comparison of the spectra of the band-to-band electroluminescence for the sample with shunting (in contrast to the sample without shunting) in the electric field with the photoluminescence spectra of the sample without shunting and at absence of the field shows absence of the spectrum maximum shift. This gives evidence that deposition of the shunting film does not impact on the emission spectrum (i.e., by essence, parameters of QW). Besides, the absence of the peak shift in the electroluminescence spectrum of the shunted sample in the electric fields as compared to the case of absence of the electric field enables us to state that a significant Joule heating of sample lattice is absent.

In order to clarify reasons for a significant difference in emission intensity we carried out measurements of its distribution along the samples either with shunting and without it at different applied fields (Fig.7). The value of emission intensity is given for the beginning of the current pulse (Fig.2).

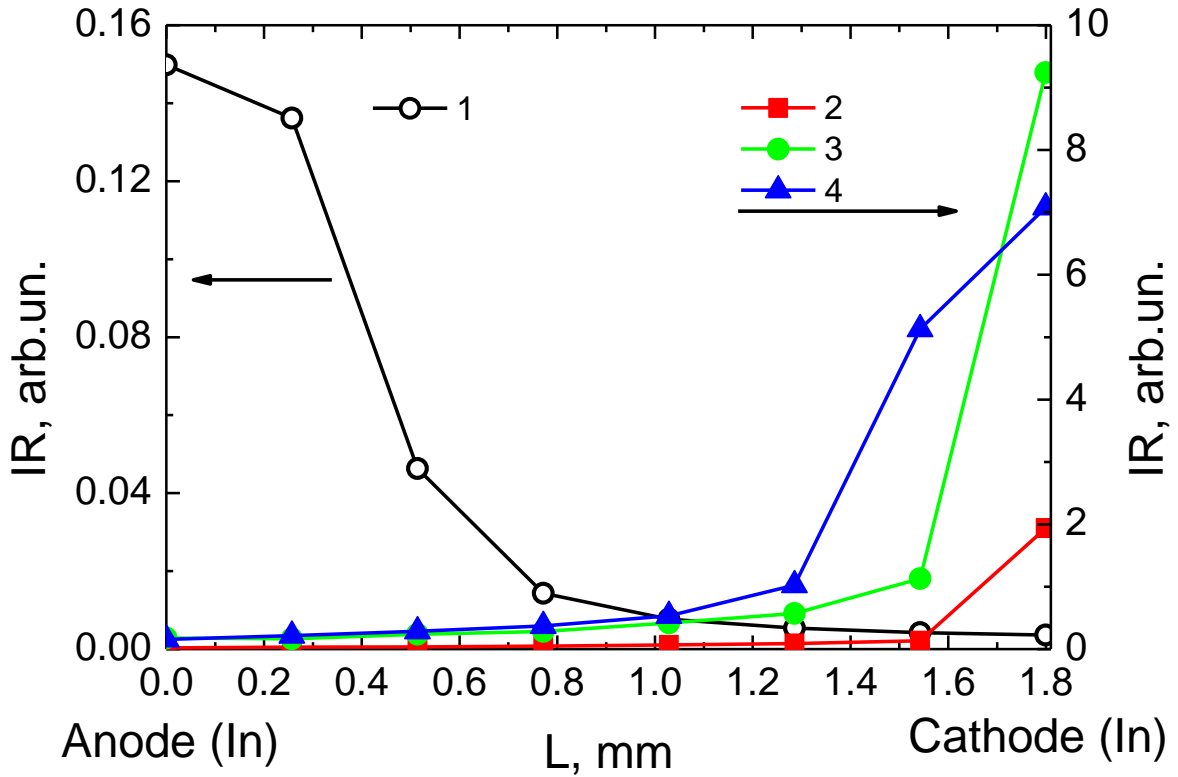

**Fig.7.** Distribution of the band-to-band emission intensity at the beginning of the current pulse along the sample at different moderate fields: 1 – without shunting (0.42 kV/cm); 2, 3, 4 – with shunting (2 - 0.2 kV/cm, 3 – 0.33 kV/cm, 4 – 0.48 kV/cm).

The distribution of the band-to-band emission intensity along the sample without shunting in the HS with double QWs was already investigated by us [7,8]. The spatial separation of electrons and holes in them occurred in a strong electric field mainly between neighbor QWs. In our case of the multi well HS with single QWs the comparatively weak emission in the moderate electric fields (about 0.5 kV/cm) in the $J_{inc}$ part is maximal near the anode contact and in fact exponentially decays in direction toward the cathode contact (Fig.7, curve 1). At that, the emission intensity decay in progress of the hole drift from the anode to the cathode is caused, aside of restriction of the hole lifetime, also by an increase of electrons part trapped into "piezo wells" of the coherent acoustic waves. Such trapping od carriers by the coherent acoustic waves is described in [19, 20]. These waves emerge near the cathode from the near-cathode acoustic noise [12]. These waves move in the direction opposite to the holes drift.

Such distribution of emission is caused by drift toward the cathode of the holes emerged from the anode contact due to interband ionization in a strong electric field [18]. The spatial separation of electrons and holes in the structures with single QWs may occur, as mentioned above, as a result of trapping electrons and holes into piezo wells of the acoustic wave. It results in increasing of the lifetime as compared to the bulk material (up to $10^{-6}$ s) and the drift length may achieve several mm [6-8].

The distribution of the band-to-band emission intensity along the sample with shunting (Fig.7, curves 2-4) differs from that for the not-shunted sample. The emission intensity near the anode of such sample in the whole range of studied fields (0.2 through 0.5 kV/cm) approximately coincides with that of the not-shunted sample. However, it now does not decay with an increasing field and smoothly increases in direction to the cathode. Closely near the contact one observes a sharp growth of the emission intensity almost by an order of magnitude. With increasing an average field this near-cathode region at first widens because of a large supply of carriers from the anode contact (Fig.7, curve 3) and then it narrows because of a stronger drag of holes toward the cathode (Fig.7, curve 4). At that the emission intensity in this region growth with the field. Such behavior of the emission distribution is caused, at first, by increasing the carrier concentration generated at the anode contact due to increasing the field at it, and, secondly, because of partly trapping of carriers into piezo wells of the acoustic wave and increasing their lifetime and length of drift filling the sample.

Usually, one uses the In ohmic contacts in experiments on the lateral transport in the heterostructures based on GaAs in strong electric fields. According to i.e. [10] they create contacts of good quality. However, the results, outlined above, indicate that already in the fields ~20 V/cm these contacts inject into the sample some quantity of the minority carriers. The cause of this, as mentioned above, is the electric breakdown on the tips of the needle-like impurity shunts in the contact [18].

Application of the Zn injecting p-n-junction as a contact does not change the qualitative picture of the acoustoelectric domain emergence, which is described above. The acoustoelectric domains with the current and emission characteristics described above also emerge in such samples. Also, the current peak, $J_{inc}$, appears in the incubation period in the strong field in the not-shunted sample, and then a sharp drop of the current and emission is observed after the domain emergence. These drops disappear in the shunted sample. Using the shunting metallic film in these samples results in suppression and elimination of the acoustoelectric domain. The main distinction consists in increasing the band-to-band emission intensity in moderate electric fields. The injection of larger concentration of carriers into the shunted sample results in a smoother distribution of the emission intensity along the sample (Fig.8).

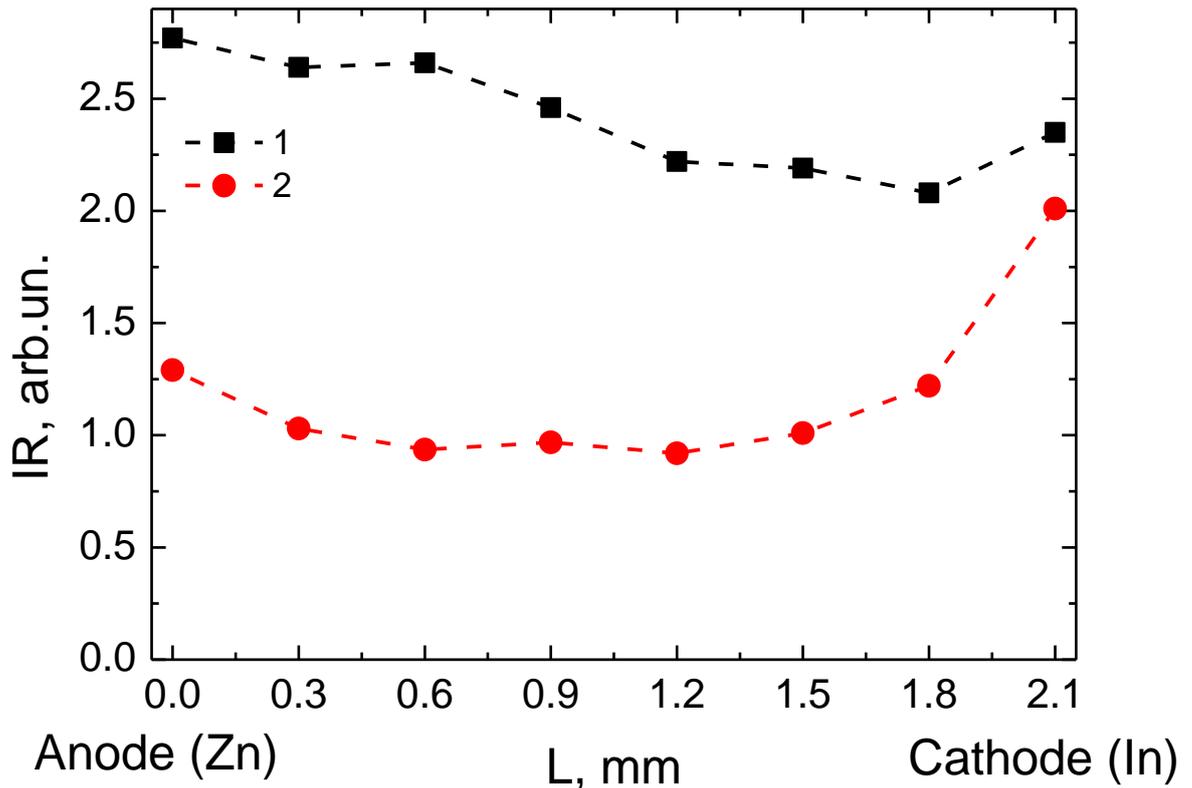

**Fig. 8.** Distribution of the band-to-band emission intensity in the shunted sample with the Zn anode injecting contact in the beginning of the current pulse at different average electric fields applied to the sample: 1- 0.58 kV/cm; 2 – 0.75 kV/cm.

A similar shunting effect with elimination of the acoustoelectric domain was obtained also for the multi-layer heterostructures GaAs/InGaAs consisting triple tunnel-coupled quantum wells and ohmic contacts, as well as for the GaAs films with suffient extent of doping ($>3·10^{11}$ cm$^{-2}$) deposited on surface of structures.

**Conclusions**

Deposition of a thin semi opaque metallic film of on the surface of the multi-layer InGaAs/GaAs heterostructure with single QWs efficiently suppresses emergence of the strong field (acoustoelectric) domain. The deposited film shunts the conduction of the sample performed by the lateral transport of charge carriers in QWs. Emergence of these domains shortens by tens times the full time of emission during the incubation period of a domain emergence (~100 through 200 ns). Besides, this frequently causes irreversible electric breakdown and damage of optoelectronic devices. Therefore, this result is interesting for applications.

Also an important effect of this shunting is alignment of the field distribution between the cathode and anode contacts. It results in an increase of the band-to-band emission in such heterostructures in comparatively small fields (~100 V/cm) by more than two orders of magnitude, even with use of ohmic contacts.

**Acknowledgment** The authors acknowledge the National Academy of Sciences of Ukraine for support (Project No. 4.8/23-p "Innovative materials and